\documentclass[twocolumn,english,prl,floatfix,citeautoscript,nofootinbib,superscriptaddress]{revtex4-2}

\usepackage{amsfonts}
\usepackage{amsbsy}
\usepackage{latexsym,epsfig,graphicx}
\usepackage{dcolumn}
\usepackage{subfigure}
\usepackage{comment}
\usepackage{color}
\usepackage[colorlinks,urlcolor=blue,citecolor=blue]{hyperref}
\usepackage{amstext}
\usepackage{amssymb}
\usepackage{setspace}

\begin{document}

\title{Self-organized Limit Cycles in Red-detuned Atom-cavity Systems}
\author{Pan Gao}
\author{Zheng-Wei Zhou}
\author{Guang-Can Guo}
\author{Xi-Wang Luo}
\email{luoxw@ustc.edu.cn}
\affiliation{CAS Key Laboratory of Quantum Information, University of Science and Technology of China, Hefei 230026, China}
\affiliation{Synergetic Innovation Center of Quantum Information and Quantum Physics,
University of Science and Technology of China, Hefei, China}
\affiliation{Hefei National Laboratory, Hefei 230088, China}

\begin{abstract}
Recent experimental advances in the field of cold-atom cavity QED provide a powerful tool for exploring non-equilibrium correlated quantum phenomena beyond conventional condensed-matter scenarios. We present the dynamical phase diagram of a driven Bose-Einstein condensate coupled with the light field of a cavity, with a transverse driving field red-detuned from atomic resonance. We identify regions in parameter space showing dynamical instabilities in the form of limit cycles, which evolve into
chaotic behavior in the strong driving limit.
Such limit cycles originate from the interplay between cavity dissipation and atom-induced resonance frequency shift, which modifies the phase of cavity mode and gives excessive negative feedback on the atomic density modulation, leading to instabilities of the superradiant scattering. We 
find interesting merging of the limit cycles related by a $Z_2$ symmetry, and identify a new type of limit cycle formed by purely atomic excitations. The effects of quantum fluctuations and atomic interactions are also investigated.
\end{abstract}

\maketitle

\section{Introduction}
In the past several decades, the interaction
between atoms and electromagnetic field of cavities
has been well studied in
both theory and experiment~\cite{RevModPhys.73.565,ritschColdAtomsCavitygenerated2013c,RevModPhys.87.1379,mivehvarCavityQEDQuantum2021,mekhovQuantumOpticsUltracold2012},
showcasing rich
cavity quantum electrodynamics (cavity-QED) physics ranging
from few-body problems such as Jaynes-Cummings model~\cite{shore1993jaynes,Fink2008} to many-body physics such as the polariton condensation~\cite{Kollar2017,lewenstein2007ultracold} and the Dicke superradiance~\cite{Dicke1954a,PhysRevLett.89.253003,nagySelforganizationBoseEinsteinCondensate2008}.
On the application side, such light-atom hybrid systems play important roles in quantum information processing~\cite{RevModPhys.87.1379,Kimble2008,PRXQuantum.2.017002,PhysRevX.8.011018}. For fundamental research, they 
provide an ideal setup for implementing and simulating solid-state Hamiltonians~\cite{mekhovQuantumOpticsUltracold2012,ritschColdAtomsCavitygenerated2013c,mivehvarCavityQEDQuantum2021,Hartmann2006,PhysRevA.76.031805,PhysRevLett.101.246809} and exploring non-equilibrium many-body phases beyond conventional condensed-matter scenarios~\cite{mekhovQuantumOpticsUltracold2012,ritschColdAtomsCavitygenerated2013c,mivehvarCavityQEDQuantum2021,chenSuperradianceDegenerateFermi2014,diehlDynamicalPhaseTransitions2010,PhysRevLett.115.045303,mivehvarSuperradiantTopologicalPeierls2017,PhysRevLett.120.263202,bakhtiariNonequilibriumPhaseTransition2015}.
A landmark example of non-equilibrium phenomena in the atom-cavity system is the Dicke superradiance, as observed experimentally with a Bose-Einstein condensate (BEC) inside a cavity, the BEC breaks translational symmetry by self-organizing onto a lattice pattern determined by the cavity mode~\cite{nagyDickeModelPhaseTransition2010c,Baumann2010a,CriticalExponentQuantumnoisedriven2011b,PhysRevLett.121.220405,baumannExploringSymmetryBreaking2011,klinderObservationSuperradiantMott2015,moralesCouplingTwoOrder2018,landiniFormationSpinTexture2018b,PNAS.Hemmerich.2015,APB.Hemmerich.2016}. Considerable experimental progress in BEC-cavity coupled systems has led to the study of various many-body problems such as long-range photon-mediated atom-atom interactions~\cite{mivehvarCavityQuantumElectrodynamicalToolboxQuantum2019a,hruby2018metastability,klinderObservationSuperradiantMott2015,landiniFormationSpinTexture2018b,kroezeSpinorSelfOrderingQuantum2018}, supersolidity and complex dynamics in multiple
cavities or in a multimode cavity~\cite{leonard2017supersolid,leonardMonitoringManipulatingHiggs2017,mivehvarDrivenDissipativeSupersolidRing2018,vaidyaTunableRangePhotonMediatedAtomic2018,ballantineMeissnerlikeEffectSynthetic2017}.

Recently, interesting dynamical instabilities in the superradiant self-organization have attracted much attention, where non-steady behavior such as the limit cycle emerges without
an explicit time-dependent external driving~\cite{keelingCollectiveDynamicsBoseEinstein2010a,kesslerEmergentLimitCycles2019b,chiacchioDissipationInducedInstabilitiesSpinor2019a,dograDissipationInducedStructural2019,piazzaSelfOrderedLimitCycles2015,Kessler2020,bucaDissipationInducedNonstationarity2019,zupancicBandInducedSelfOrganization2019,linPathwayChaosHierarchical2020a}, which shares strong similarities with time crystals~\cite{Sacha2017,PhysRevLett.109.160401}. To obtain stable limit cycles, prior studies either employ the blue-detuned BEC-cavity system~\cite{keelingCollectiveDynamicsBoseEinstein2010a,kesslerEmergentLimitCycles2019b,piazzaSelfOrderedLimitCycles2015,Kessler2020,linPathwayChaosHierarchical2020a}, or utilize the spinor BECs with competing density and spin couplings with cavity mode~\cite{chiacchioDissipationInducedInstabilitiesSpinor2019a,dograDissipationInducedStructural2019,bucaDissipationInducedNonstationarity2019}. For the blue-detuned BEC-cavity system, the limit cycle results from the interplay between collective coherent scattering
and low-field dragging dipole force~\cite{piazzaSelfOrderedLimitCycles2015}; for the spinor-BEC-cavity system, it results from the cavity field mediated nonreciprocal coupling between the two collective spins, which is due to the competing density- and spin-wave scattering together with the dissipation-induced phase shift of the cavity mode~\cite{chiacchioDissipationInducedInstabilitiesSpinor2019a}. In contrast, for the experimentally more accessible red-detuned single-component BEC-cavity system, most previous studies have focused on the stable steady-state superradiance~\cite{mivehvarCavityQEDQuantum2021}, the instability properties are not well-explored.

In this paper, we investigate dynamical instabilities 
of a BEC inside a high finesse
optical cavity, with a transverse 
driving field red-detuned from the 
relevant atomic resonance.
We map out the dynamical phase diagram and uncover novel instabilities 
induced by the interplay between cavity dissipation and atom-induced resonance shift.
By increasing the pump rate, the system first undergoes a transition from normal phase to superradiant phase, spontaneous symmetry breaking takes place between two possible stable steady states related by a $Z_2$ symmetry~\cite{nagyDickeModelPhaseTransition2010c,CriticalExponentQuantumnoisedriven2011b,baumannExploringSymmetryBreaking2011}. Then the system enters the unstable region where each steady state evolves into a superradiant limit cycle that spontaneously breaks the time translation symmetry.
Interestingly, we find that before the system enters the chaotic region in the strong driving limit, the two limit cycles (related by the $Z_2$ symmetry) may first merge together as the pump rate increases, leading to a single limit cycle and restoring the $Z_2$ symmetry. Moreover, we identify a new limit-cycle phenomenon with purely atomic excitation, the cavity field is suppressed to zero by the interference between scatterings from different momentum states.
In contrast to previous works~\cite{keelingCollectiveDynamicsBoseEinstein2010a,kesslerEmergentLimitCycles2019b,piazzaSelfOrderedLimitCycles2015,Kessler2020,linPathwayChaosHierarchical2020a,chiacchioDissipationInducedInstabilitiesSpinor2019a,dograDissipationInducedStructural2019,bucaDissipationInducedNonstationarity2019},
here 
the atom-induced frequency shift of the cavity resonance
gives excessive negative feedback on the atomic density modulation through shifting the phase of cavity mode, which is responsible for the instability. 
We find that the limit cycles with purely atomic excitation are not affected by quantum fluctuations.
While for the superradiant limit cycles, the time-domain oscillations of order parameters averaged over stochastic trajectories are suppressed, but the frequency-domain peaks persist.
All the dynamical phases mentioned above survive in the presence of atom-atom interaction, though the phase boundaries may be modified accordingly.

\begin{figure}[t]
\includegraphics[width=\linewidth]{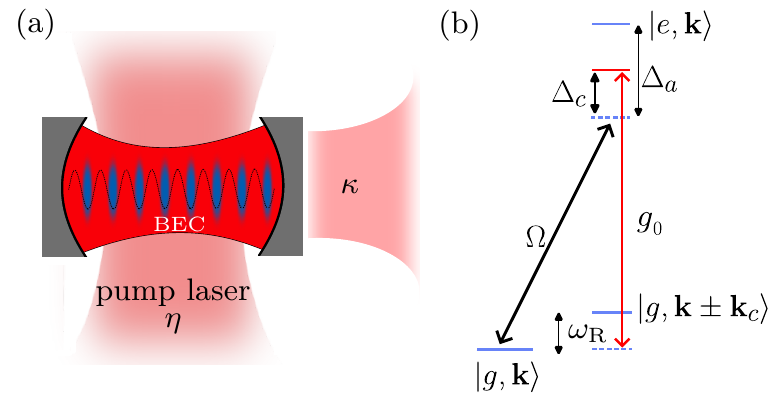}
\caption{(a) Schematic of the proposed experimental setup. A BEC trapped inside an optical cavity is transversely driven by a pump laser.
(b) Energy levels of the atoms and couplings induced by the red-detuned pump and cavity fields with Rabi frequencies $\Omega$ and $g_0$ respectively.}
\label{fig:setup}
\end{figure}

\section{Model}
We consider a gas of 
ultracold atoms forming a BEC trapped inside a high-finesse optical cavity,
as sketched in Fig.~\ref{fig:setup}a. The BEC is transversely pumped by a coherent light that is red detuned to an atomic transition, the detuning $\Delta_a$ is large and the excited atomic level can be eliminated adiabatically. Therefore,
the BEC couples with the single mode of the cavity through
a two-photon Raman scattering process between
the cavity and the driving fields (see Fig.~\ref{fig:setup}b),  
accompanied by transitions between
the BEC ground state $|g,\mathbf{k}\rangle$ and the excited momentum states $|g,\mathbf{k}+\mathbf{k}_c\rangle$, with $\mathbf{k}_c$ the wave vector of the cavity mode. The dynamics of such
driven-dissipative atom-cavity systems is well described by the following coupled equations of motion~\cite{nagySelforganizationBoseEinsteinCondensate2008, piazzaSelfOrderedLimitCycles2015}
\begin{eqnarray}
  i \hbar \partial_t \Psi(x, t) &=&\bigg[-\frac{\hbar^2 \partial_{x x}^2}{2 m}+g_\text{aa}|\Psi|^2+\hbar U_0|\alpha|^2 \cos ^2(k_c x) \nonumber \\
  & & + \frac{\hbar\eta} {\sqrt{N}} (\alpha+\alpha^*) \cos (k_c x)\bigg] \Psi(x, t) \nonumber \\
  i \partial_t \alpha &=&\left[\Delta_{c}-i \kappa +U_{0} N \mathcal{B}\right] \alpha+\eta\sqrt{N}\Theta + i\xi
  \label{eq:eom}
\end{eqnarray}
where $\Psi(x,t)$ is the BEC wave function and $\alpha$ is the expectation value of the cavity field, we restrict the motion of the atoms
along the cavity axis $x$ by assuming additional trapping in the other
directions. $N$ is the atom number, $g_\text{aa}$ is the atom-atom interaction strength, and $U_0=-\frac{g_0^2}{\Delta_a}$ is the AC Stark shift induced by a single photon
as well as the frequency shift of
the cavity resonance induced by a single atom (at antinodes). $\eta=\frac{\sqrt{N}g_0\Omega}{\Delta_a}$ is the effective pump rate, $\Delta_c$ is the detuning of the cavity mode (see Fig.~\ref{fig:setup}b), $\kappa$ is the cavity dissipative rate. For our red-detuned system, one has $U_0<0$ and $\Delta_c>0$. $\Theta=\int \rho(x,t) \cos(k_cx)$ and $\mathcal{B}=\int \rho(x,t) \cos^2(k_cx)$ are the atomic order parameters associated with the superradiance, with  $\rho(x,t)=|\Psi|^2/N$ the normalized atomic density. To take into account the effects of quantum fluctuation,
we included in Eq.~\ref{eq:eom}
the stochastic noise term $\xi(t)$ associated with cavity dissipation~\cite{ritschColdAtomsCavitygenerated2013c}, for the mean-field (MF) solutions, we simply drop the
$\xi$ term. In the following, we will first focus on the MF results, and discuss the quantum fluctuation effects later.

Notice that
the system possesses a $Z_2$ symmetry, associated with invariance under the transformation of cavity field and BEC wave function such that
$\{\alpha,\Theta\}\rightarrow\{-\alpha,-\Theta\}$~\cite{nagyDickeModelPhaseTransition2010c,CriticalExponentQuantumnoisedriven2011b,baumannExploringSymmetryBreaking2011}.
In the following numerical simulation, we will consider the experimentally realistic $^{87}$Rb BEC~\cite{PNAS.Hemmerich.2015,APB.Hemmerich.2016,PhysRevLett.121.220405} 
with atom number $N=10^5$ and recoil frequency $\omega_R=\hbar k_c^2/2m\simeq 2\pi \times 3.7$ kHz. Both the cavity detuning $\Delta_c$ and dissipative rate are  on the order of tens of kHz, and we set $\kappa=10\omega_R$ all through the paper for simplicity.
For the cavity resonance, we consider a strong atom-induced shift
$|U_0|N\gtrsim \Delta_c$, which is crucial for the emergency of limit cycles. 

\begin{figure}[tb]
\includegraphics[width=1.0 \linewidth]{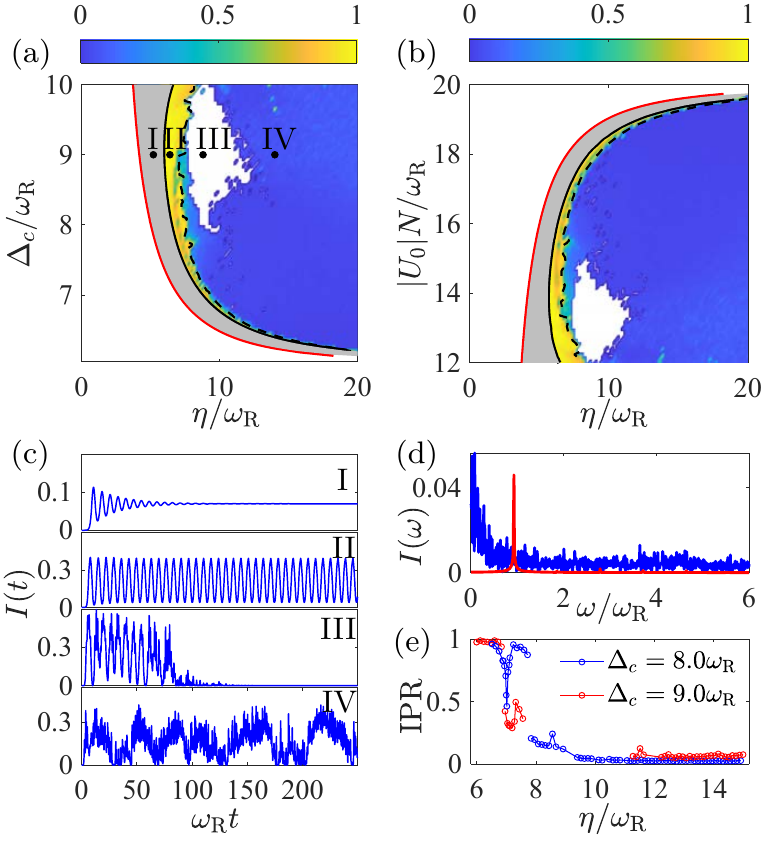}
\caption{(a) and (b) Dynamics phase diagrams in the $\Delta_{c}$-$\eta$  and $|U_0|N$-$\eta$ planes,
with $U_{0} N=-12\omega_{\rm{R}}$ and $\Delta_c=10\omega_{R}$, respectively. The red solid line denotes the N (left white area) to S (gray area) phase boundary, the black solid line gives the boundary between S and SL phase. The color bars represent the IPR of cavity field intensity in the frequency domain, with sudden drops across the black dotted lines which signal the transition from SL to C phase. The right white area corresponds to AL phase.
(c) The time evolution of cavity field in different phases, the parameters in panels I, II, III and IV are $\eta=5.2\omega_{\rm{R}},6.4\omega_{\rm{R}},8.8\omega_{\rm{R}},14\omega_{\rm{R}}$ and $\Delta_{c}=9.0\omega_{\rm{R}}$, as marked by the dots in (a).  (d) Typical frequency distributions of cavity-field intensity in the SL (red line) and C (blue line) phases, with parameters given by dots
$\rm{\uppercase\expandafter{\romannumeral2}}$ and 
$\rm{\uppercase\expandafter{\romannumeral4}}$ in (a).
(e) Spectrum IPR of cavity field intensity along the lines $\Delta_{c}=8.0\omega_{\rm{R}}$ (blue) and $\Delta_{c}=9.0\omega_{\rm{R}}$ (red) in (a). In all plots, we have $g_\text{aa}=0$.}
\label{fig:phase_diagram}
\end{figure}

\section{Phase diagram}
We present in Figs.~\ref{fig:phase_diagram}a and \ref{fig:phase_diagram}b the MF dynamical phase diagrams in the $\eta$-$\Delta_c$ and $\eta$-$|U_0|N$ planes, respectively, which are obtained by solving
Eq.~\ref{eq:eom} and analyzing their long-time behaviors. 
We choose a homogeneous BEC $\Psi=\sqrt{N/L}$ as
initial conditions with $L$ the system size 
and an infinitesimally occupied cavity $\alpha(0)/\sqrt{N}\ll 1$ as a seed. We will consider a fixed pump strength during the evolution, we have verified that the dynamical phenomena discussed in this paper are not affected by considering an initial ramping protocol for the pump strength.
We find five different dynamical phases: 
(N) normal phase, with vanishing cavity field and homogeneous BEC as the stable steady state; (S)  superradiant phase, with self-organized striped BEC and finite cavity field as the stable steady  state; (SL) superradiant limit-cycle phase without a stable steady state, where both the superradiant cavity field and the density pattern of the BEC  
develop into periodic self-sustaining  oscillations; (AL) atomic limit-cycle phase with only atomic excitations and vanishing cavity field, where the stripe pattern of the BEC oscillates periodically due to interference between different momentum states; (C) chaotic phase with irregularly oscillating order parameters $\alpha,\mathcal{B}, \Theta$.	
In Fig.~\ref{fig:phase_diagram}c (from the top panel to the bottom panel), we plot the typical time
evolution of the renormalized cavity mode intensity $I(t)=|\alpha|^2/N$ for the S, SL, AL and C phases.

In the weak pumping region, the system is in a normal phase (the white areas on the left of Figs.~\ref{fig:phase_diagram}a and \ref{fig:phase_diagram}b), and undergoes a transition to the superraidiant phase (gray areas) as we increase the pump rate. At the vicinity of the N-S phase boundary, the cavity field $\alpha$ is weak and the $\eta \cos(k_cx)$ term dominates the dynamics, where the system is characterized by the conventional 
two-mode superradiant physics, and
our numerical critical pump rate (red solid lines in Figs.~\ref{fig:phase_diagram}a and \ref{fig:phase_diagram}b) is perfectly in agreement with the analytical result
$\hbar^2\eta_c^2=(\hbar\omega_R/2+g_\text{aa}N/L)(\delta_0+\kappa^2/\delta_0)$~\cite{nagySelforganizationBoseEinsteinCondensate2008} with
$\delta_0=\Delta_c+U_0N/2$. 
We see that the transition towards a stable superradiant phase
occurs only for $\delta_0>0$ (i.e., $\Delta_c>|U_0|N/2$).
As we further increase the pump rate, the long-time superradiant cavity field $|\alpha|^2$ and atomic order parameters $\Theta$, $\mathcal{B}$ become stronger, the effect of the $U_0$ term in Eq.~\ref{eq:eom} becomes significant. 
Here the cavity has a strong decay rate, for our analytical considerations, we can eliminate the cavity field by assuming that its value follows $\Theta$ adiabatically as $\alpha=\frac{\eta \sqrt{N}\Theta}{\delta_\text{eff}-i\kappa}$, with $\delta_\text{eff}=
\Delta_c+U_0N\mathcal{B}$ the effective cavity detuning which can alter the phase of $\alpha$ significantly. 
On one hand, a larger $\Theta$ leads to a stronger cavity field $|\alpha|$ and thereby a larger AC-stark potential depth $U_0|\alpha|^2$, 
which gives a stronger $\mathcal{B}$. On the other hand,
a larger $\mathcal{B}$  would reduce $\text{Re}[\alpha]$ through changing the cavity resonance (i.e., $\delta_\text{eff}$), leading to a weaker Raman scattering potential $\frac{\hbar\eta} {\sqrt{N}} (\alpha+\alpha^*) \cos (k_c x)$ which tends to reduce  $\Theta$.
Such negative feedback between $\mathcal{B}$ and $\Theta$ becomes excessive such that the system cannot find a stable steady state beyond some critical pump rate (black solid lines in Figs.~\ref{fig:phase_diagram}a and \ref{fig:phase_diagram}b). 
The dynamical instability can also be seen by looking at the imaginary part of the
eigenvalue of the maximally growing mode resulting from the linear stability analysis of the
steady-state solution (see Appendix for more details)~\cite{piazzaSelfOrderedLimitCycles2015,nagySelforganizationBoseEinsteinCondensate2008},
and we find that
the maximal growth rate changes from zero to positive across the above critical pump rate.
Such instability does not simply lead to heating and collapse of the order. 
Instead, the BEC starts to periodically oscillate between different ordered patterns (together with an oscillating cavity field),
and the system enters the SL phase. The oscillation frequency is on the order of $\omega_R$.

In the strong pump limit, the limit cycles turn
into chaotic dynamics as the BEC and cavity field oscillate irregularly with indefinite number of frequencies.
We expand the wave function as
$\Psi=\sqrt{N/L}\sum_n c_n e^{ink_cx}$ (with integer $n$) and find that a strong pump rate leads to the macroscopic populations of the BEC on high momentum states. Nevertheless, the momentum distribution is still well localized around $n=0$
since the cooling effects of the cavity dissipation prevents unbounded increase of the kinetic
energy~\cite{piazzaSelfOrderedLimitCycles2015}. Notice that the scattering of the pump light into the cavity is associated with the coupling between adjacent momentum states of the BEC, when high momentum states are populated, the destructive interference of scattered light from left and right momentum neighbors 
may suppress such scattering process. A new type of attractor is developed where the cavity field vanishes at long time while the BEC occupies only the momentum states with even $n$ (i.e., $n=0,2,\cdots$), as indicated by the white areas (where the long-time intensity $|\alpha|^2$ averaged over $\omega_R t\in[1500,2000]$ is less than $10^{-3}$) at the right unstable regions in Figs.~\ref{fig:phase_diagram}a and \ref{fig:phase_diagram}b. The superposition of different momentum states leads to the  periodic self-sustained oscillations between different ordered patterns of the BEC, with dominant oscillating frequency $4\omega_R$ (due to interference between $n=0$ and $n=2$), and we denote such oscillations as the atomic limit cycles. The AL phase possesses the $Z_2$ symmetry due to the vanishing cavity field.

To identify the boundary between the superradiant limit-cycle (SL) and chaotic (C) dynamics, we examine the cavity-field intensity in the frequency domain
$I(\omega)=\int  I(t)e^{-i\omega t}dt$ and define the inverse participation ratio (IPR) as
\begin{equation}
    \text{IPR}=\frac{\int  |I(\omega)|^4 d\omega}{\int |I(\omega)|^2 d\omega }
    \label{eq:IPR}
\end{equation}
which measures the locality~\cite{RevModPhys.80.1355} of the frequency distribution. For limit cycles, $I(\omega)$ involves several well
defined frequency components (with one dominant) as shown in Fig.~\ref{fig:phase_diagram}d, and thus the SL phase has a high IPR. For the chaos,  $I(\omega)$ involves indefinite number of frequencies (see Fig.~\ref{fig:phase_diagram}d), leading to  a low IPR. We find sharp jumps of the IPR from the value close to 1 to the value well below 0.5 (see Fig.~\ref{fig:phase_diagram}e), which marks the boundary between SL and C phases. Note that the IPR defined in Eq.~\ref{eq:IPR} is not applicable in the AL phase. The color bars in Figs.~\ref{fig:phase_diagram}a and \ref{fig:phase_diagram}b denote the corresponding IPR and the dashed lines separate the SL and C phases. In Fig.~\ref{fig:phase_diagram}e, we note that the IPR for $\Delta_c=8\omega_R$ has a dip within the SL phase, which is due to the merging of the limit-cycle pairs related by the $Z_2$-symmetry, as we shall discuss in more details later.

\begin{figure}[tb]
\includegraphics[width=1.0 \linewidth]{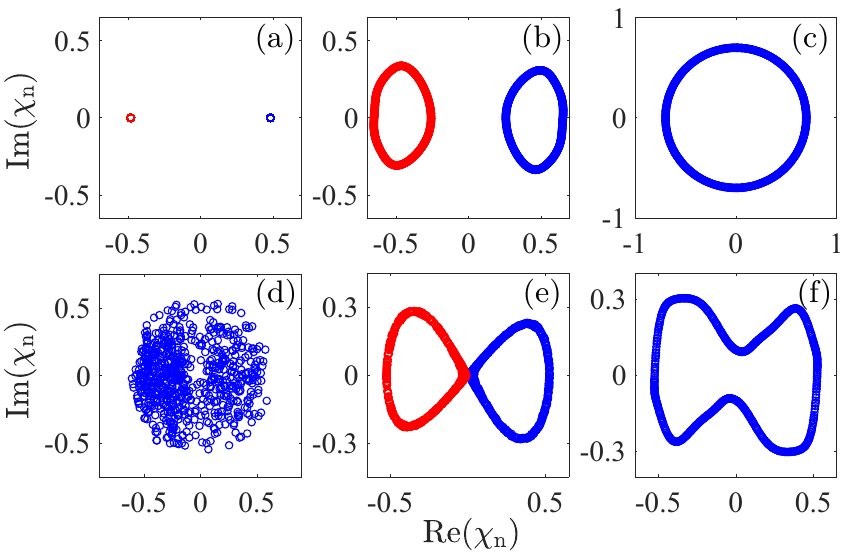}
\caption{(a)-(d) Orbits of the atomic correlation $\chi_n$ in the S, SL, AL and C phases,  with parameters marked by the dots I, II, III and IV in Fig.~\ref{fig:phase_diagram}a, respectively. The two orbits related by the $Z_2$ symmetry are shown in red and blue.
(e) Merging of two limit cycles related by the $Z_2$ symmetry with $\eta=7\omega_{\rm{R}}$ and $\Delta_c=8\omega_R$, corresponding to the dip of the IPR in Fig.~\ref{fig:phase_diagram}e.
(f) $Z_2$ symmetric limit cycle after the merging with $\eta=7.2\omega_{\rm{R}}$ and $\Delta_c=8\omega_R$. We have used $n=2$ for AL phase in (c) and $n=1$ for other sub-figures. Other parameters are the same as that in Fig.~\ref{fig:phase_diagram}a.}
\label{fig:BEC_orbits}
\end{figure}

Besides the cavity field, the oscillations and recurrences can also be seen in the
condensate wave function for the limit cycles. We consider the momentum-space correlations
$\chi_{n}=c_0 (c_n^*+c_{-n}^*)$ and plot their orbits in the $\text{Re}[\chi_n]$-$\text{Im}[\chi_n]$ plane in Fig.~\ref{fig:BEC_orbits}.
The orbits reduce to fix points in the stable superradiant phase (S), as shown in Fig.~\ref{fig:BEC_orbits}a. There are two fixed points corresponding to the two steady-state solutions related by the $Z_2$ symmetry. Each fixed point involves into a periodic orbit in the SL phase, as shown in Fig.~\ref{fig:BEC_orbits}b. Fig.~\ref{fig:BEC_orbits}c shows the periodic orbit for $\chi_2$ in the AL phase where only the even recoil momenta are populated, while Fig.~\ref{fig:BEC_orbits}d shows the irregular orbit of $\chi_1$ in the chaotic phase. As we mentioned before,
the SL phase may merge the limit cycles related by a $Z_2$ symmetry before evolving
into C phase, the periodic orbits (i.e., limit cycles) in Fig.~\ref{fig:BEC_orbits}b become larger and larger as the pump rate increases. At some critical pump, the two orbits related by the $Z_2$ symmetry may close their gap at the origin and merge into a single $Z_2$-symmetric orbit,
which can be clearly seen from Figs.~\ref{fig:BEC_orbits}e and \ref{fig:BEC_orbits}f.
At the merging point, the system can switch between the two orbits and the dynamics losses the periodicity, leading to the drop of the IPR as observed in Fig.~\ref{fig:phase_diagram}e.
Depending on the value of $\Delta_c$ and $U_0N$, 
a single but larger limit cycle may be generated after the merging (see Fig.~\ref{fig:BEC_orbits}f), and further increasing the pump rate would drive the dynamics into chaotic behavior, as shown in Fig.~\ref{fig:phase_diagram}e with $\Delta_c=8\omega_R$. Right after the merging, the periods of the oscillations remain the same for $|\alpha|^2$, but are doubled for $\Theta$, $\alpha$ and $\chi_1$ whose orbits are twice larger.
Also, the system may directly enter the chaotic phase at the merging point, as shown in Fig.~\ref{fig:phase_diagram}e with $\Delta_c=9\omega_R$ (after the merging, the IPR drops to about 0.4 corresponding to weak chaotic motion). For even larger $\Delta_c$, the SL phase may change into the C or AL phases before the merging.

\begin{figure*}[tb]
\includegraphics[width=1.0 \linewidth]{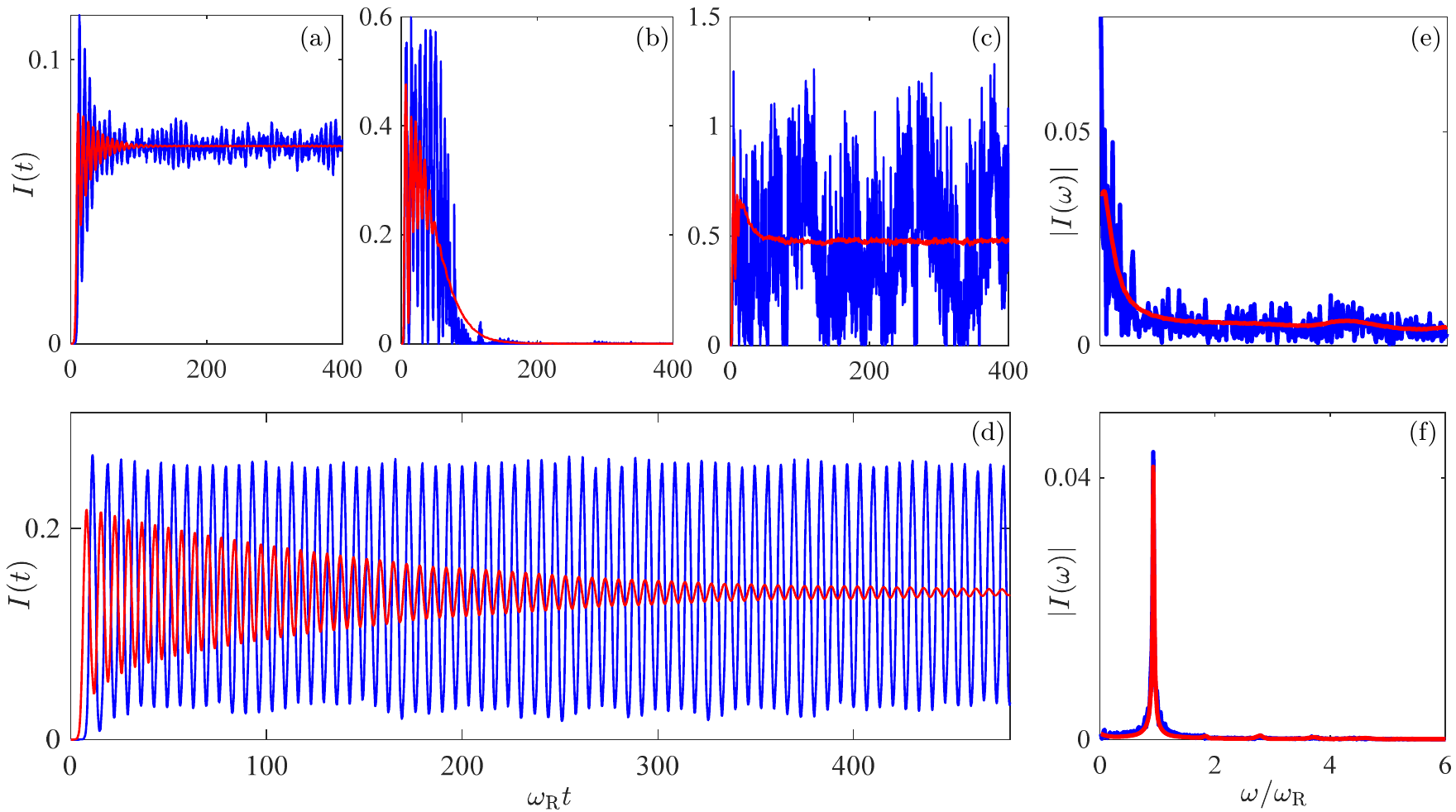}
\caption{The effects of quantum fluctuations on the dynamics according to TWA. The cavity fields in the S, AL, C and SL phases are shown in (a), (b), (c) and (d), with $\Delta_c$ and $\eta$ marked by the dots I, II, IV and III in Fig.~\ref{fig:phase_diagram}a, respectively.
The corresponding frequency distributions in the C and SL phases are shown in (e) and (f). Red lines are results averaged over $6\times 10^3$ trajectories based on TWA, blue lines are the results from a single trajectory. We verified that the TWA oscillation amplitude in (c) decays to zero eventually at long times.
Other parameters are the same as that in Fig.~\ref{fig:phase_diagram}a.}
\label{fig:TWA}
\end{figure*}

\begin{figure*}[tb]
\includegraphics[width=1.0 \linewidth]{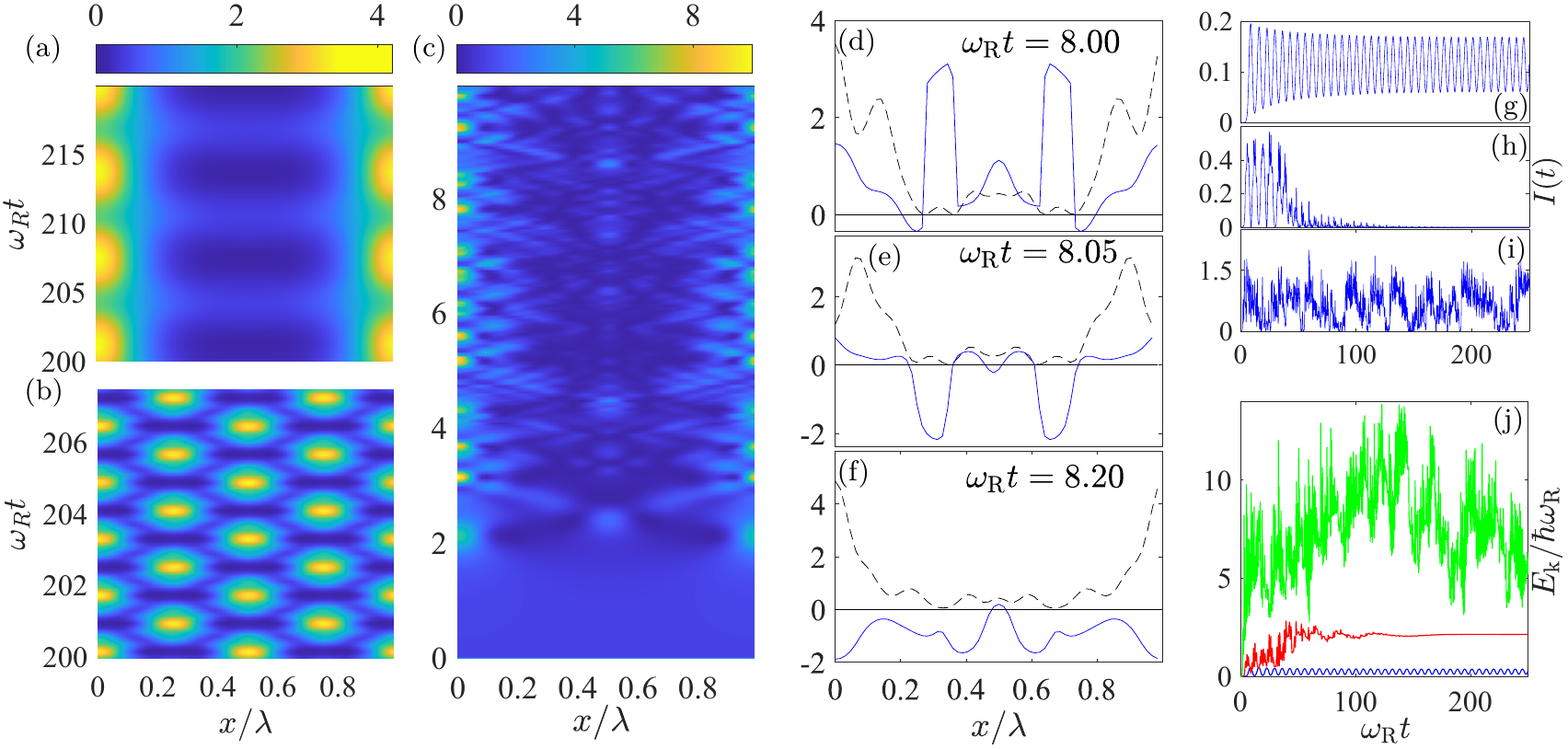}
\caption{The evolution of the BEC wave function and the cavity field in the presence of atom-atom interactions. Thanks to the 
cavity dissipation that compensates the heating due to the dynamical
instabilities, the short-range atom-atom interactions do not lead to proliferation of phase singularities (zero-density points) in the oscillating BEC. The periodic oscillations of the BEC density (color bars in units of $2\pi N/L$)
in the SL and AL phases are clearly seen in (a) and (b), with $\eta=6\omega_R$ and $\eta=8.8\omega_R$, respectively. Irregular evolution is observed in (c) for the C phase with $\eta=17.6\omega_R$.
(d)-(f) The creation and annihilation of phase singularities (i.e., $\pi$-phase jumps of $\Psi(x,t)$ and density zeros of $|\Psi(x,t)|^2$) during the evolution shown in (c), solid (dashed) lines are the phases (density) of the condensate.
(g)-(i) The evolution of cavity field corresponding to (a)-(c).
(j) The kinetic energy of the BEC, blue, red and green lines correspond to the evolution of the SL, AL and C phases shown in (a), (b) and (c) respectively.
Common parameters are: $\Delta_c=8.8\omega_R$,  $U_{0}N =-12\omega_{\rm{R}}$ and $g_\text{aa}N/L  =0.5\hbar\omega_R$.}
\label{fig:interaction}
\end{figure*}

\section{quantum fluctuation Effects}
In order to investigate the robustness of the dynamical phases against imperfections such as the quantum fluctuation, we adopt truncated Wigner approximation (TWA)~\cite{polkovnikovPhaseSpaceRepresentation2010a,Blakie.2008} 
as used in~\cite{kesslerEmergentLimitCycles2019b}.
The TWA simulates the dynamics of quantum fields 
by treating quantum operators as classical numbers and then solving classical equations
of motion, with initial conditions drawn from the quantum
Wigner distribution, and the Langevin quantum fluctuation operator is replaced by classical stochastic noise term $\xi(t)$
satisfying $\langle \xi^*(t)\xi(t')\rangle=\kappa\delta(t-t')$.  Accurate results can be obtained by the TWA when the particle number $N$ is large and the two-body interaction $g_\text{aa}$ is weak.
In our simulations, we sample the initial state by including the quantum noise of the BEC in zero-momentum mode as well as the vacuum fluctuations of the high-momentum atomic modes and the cavity mode.

The time evolution of the cavity field in different phases is shown in Figs.~\ref{fig:TWA}a-d, the results with a single-trajectory 
as well as averaged over $6\times10^3$ trajectories are both shown.
For the S phase (see Figs.~\ref{fig:TWA}a), the TWA predicts a finite
and stable cavity field with long-time expectation values in good agreement with the MF results, fluctuations on top of the mean value are observed for a single trajectory.
For the AL phase (see Figs.~\ref{fig:TWA}b), the TWA and MF results are also in good agreement, in fact, every trajectory will end up with vanishing cavity field, though the short-time evolution at the beginning may vary from one trajectory to another, indicating that the AL phase is robust against quantum fluctuations.
However, for the C and SL phases, the TWA results are very different from the MF results (see Figs.~\ref{fig:TWA}c and \ref{fig:TWA}d). The average over an ensemble of
chaotic trajectories in TWA results in a cavity field dynamics with smaller temporal fluctuations (see Figs.~\ref{fig:TWA}c), meanwhile, the fluctuations in the frequency domain are also suppressed (see Fig.~\ref{fig:TWA}e).
Unfortunately, the emergent temporal oscillations in the SL phase
decay with time for the TWA results, which means that the long-time temporal coherence of the limit cycles is lost due to the quantum fluctuations; nevertheless, we note that the system exhibits clear quasiperiodic oscillations in each trajectory, and thus there must exist a sharp peak in the frequency domain (see Figs.~\ref{fig:TWA}c and \ref{fig:TWA}f) even after the TWA average. Different from the MF results, here the TWA frequency peak is slightly broadened by the quantum fluctuations, the width of the peak broadening in the frequency domain is inversely related with the temporal width of the oscillating envelope.

Moreover, we find that the effects of quantum fluctuations is dominated by the stochastic noise term $\xi(t)$, the fluctuations in the initial state only slightly shift the phase of the oscillations, they do not affect the periods of the limit cycles in SL phase. Therefore, in the absence of the stochastic noise term $\xi(t)$, 
long-time temporal coherence of the superradiant limit cycles
would persist even in the presence of quantum fluctuations in the initial state.

\section{Interaction Effects}
In this section, we study the effects of atom-atom interactions on the dynamical phases. Our numerical simulations show that all the five dynamical phases survive in the presence of atom-atom interactions, though the phase boundaries may be modified.
The time evolutions of BEC wave function and the cavity field are illustrated in Fig.~\ref{fig:interaction} with $g_\text{aa}N/L=0.5\hbar\omega_R$, besides, periodic orbits similar to that shown in Fig.~\ref{fig:BEC_orbits} are also observed. The red-detuned atom-cavity system shows similar phase slippage dynamics as that obtained in the blue-detuned system~\cite{piazzaSelfOrderedLimitCycles2015}.
In the presence of atom-atom interaction,
the oscillating BEC may create phase singularities to lower the kinetic energy due to its superfluid nature~\cite{wrightDrivingPhaseSlips2013}. 
Such phase slips occur
periodically in the SL and AL phases (see Figs.~\ref{fig:interaction}a and \ref{fig:interaction}b), and appear
irregularly with a faster rate in the C phase (see Fig.~\ref{fig:interaction}c).
As discussed in~\cite{piazzaSelfOrderedLimitCycles2015}, the cavity dissipation counteracts this phase slip process (see Figs.~\ref{fig:interaction}d-\ref{fig:interaction}f) by subtracting energy
from the system, which prevents phase slip proliferation and ensures 
the validity of the MF approach. That is, the cavity cooling
compensates the heating due to the dynamical
instabilities.
The corresponding cavity-field evolution is shown in Figs.~\ref{fig:interaction}g-\ref{fig:interaction}i for different phases, which shows similar behavior as the non-interacting $g_\text{aa}=0$ case.
In all phases, the BEC is well localized near $n=0$ in the momentum space, therefore the kinetic energy fluctuates around a finite value after an initial increase, as shown in Fig.~\ref{fig:interaction}j, and a larger pump rate generally leads to a broader momentum-space distribution and thereby a larger kinetic energy of the BEC.

\section{Discussion and Conclusion}
We emphasize that in contrast to previously blue-detuned~\cite{piazzaSelfOrderedLimitCycles2015} or spinor-BEC systems~\cite{chiacchioDissipationInducedInstabilitiesSpinor2019a}, here the strong atom-induced cavity resonance shift (i.e., $|U_0|N\gtrsim \Delta_c$) is essential for the emergency of instabilities and limit cycles. Such resonance shift would modify the phase of cavity field through the dissipation process, which may lead to excessive negative feedback on the atomic density modulation, making the system unstable. 
A semi-classical analysis based on an atomic gas inside a ring cavity had 
predicted frustration phenomena in the
region $|U_0N|\gtrsim \Delta_c$~\cite{Nagy2006a}. By investigating the BEC-cavity dynamics in such  strong $U_0$ region with a red-detuned transverse pump, we show that not only stable superradiance, but also interesting limit cycles can be generated at sufficient pump strength. 
In the future, it would be interesting to investigate the effects of realistic harmonic traps, the transverse dynamics as well as atomic correlations in the system.

In summary, we have presented the dynamical phase diagrams of a red-detuned atom-cavity system. We identified regions with stable superradiance as well as dynamical instabilities in the form of limit cycles which evolve into chaotic behavior in strong pump limit. We predicted two types of limit cycles, one is characterized by self-sustained oscillations of both the condensate density and cavity field, the other is for condensate oscillation only. We find interesting merging of limit cycles (related by the $Z_2$ symmetry), which leads to a single $Z_2$ symmetric limit cycle. As shown in Figs.~\ref{fig:phase_diagram}a and \ref{fig:phase_diagram}b, the limit cycles can emerge in a wide range of parameters $\Delta_{c}$ and $|U_0|N$, and they are very robust against quantum fluctuations and atom-atom interactions (though for the SL phase, the time-domain coherence is lost due to quantum fluctuation, the sharp peak in frequency domain persists). 
Our work paves the way for exploring novel 
many-body limit cycle dynamics and
provides a feasible scheme for ongoing research on 
dissipative time crystals.

\section{acknowledgments}
This work is supported by Innovation Program for Quantum Science and Technology (Grant No. 2021ZD0301200) and the USTC start-up funding.


\setcounter{figure}{0} \renewcommand{\thefigure}{A\arabic{figure}} %
\setcounter{equation}{0} \renewcommand{\theequation}{A\arabic{equation}}

\section{Appendix}
\textbf{\emph{Steady state and stability.---}}Here we show how to solve for the steady state (stationary point) of Eq.~\ref{eq:eom} and determine the stability of 
these solutions by
linearizing Eq.~\ref{eq:eom} around the steady state. For simplicity, we will set $g_\text{aa}=0$ in the following.
We first rewrite the equation of motion as
\begin{eqnarray}
  i \partial_t \Psi(x, t) &=&\bigg[-\frac{\hbar \partial_{x x}^2}{2 m}+ U_0|\alpha|^2 \cos ^2(k_c x)  \nonumber \\  
  & & + \frac{\eta} {\sqrt{N}} (\alpha+\alpha^*) \cos (k_c x)\bigg] \Psi(x, t) \label{eq:eomphi} \\
  i \partial_t \alpha &=&\left[\Delta_{c}-i \kappa +U_{0} N \mathcal{B}\right] \alpha+\eta\sqrt{N}\Theta.
  \label{eq:eomA}
\end{eqnarray}
The steady state satisfies $ i \partial_t \alpha=0$ which leads to 
\begin{equation}
    \alpha=\frac{\eta \sqrt{N}\Theta}{\delta_\text{eff}-i\kappa}
    \label{eq:alphaA}
\end{equation}
with $\delta_\text{eff}=
\Delta_c+U_0N\mathcal{B}$. Then by substituting Eq.~\ref{eq:alphaA} into Eq.~\ref{eq:eomphi}, we arrive at the dynamical equation of $\Psi(x,t)$ with effective long-range nonlinear interactions.
The steady state solution of $\Psi(x,t)$ satisfies
$i  \partial_t \Psi(x, t)=\mu \Psi(x, t)$ with $\mu$ the chemical potential. To obtain the steady state solution,
we use a variant of the imaginary time evolution method, which 
consists of propagating $\Psi(x,t)$ in imaginary
time $\tau = it$ according to  Eq.~\ref{eq:eomphi}, with 
cavity field replaced by $\alpha=\frac{\eta \sqrt{N}\Theta}{\delta_\text{eff}-i\kappa}$. 
The N-S phase transition can also be determined by examining the steady state cavity field $\alpha$ which changes from zero to a finite value.

\begin{figure}[tb]
\includegraphics[width=1.0 \linewidth]{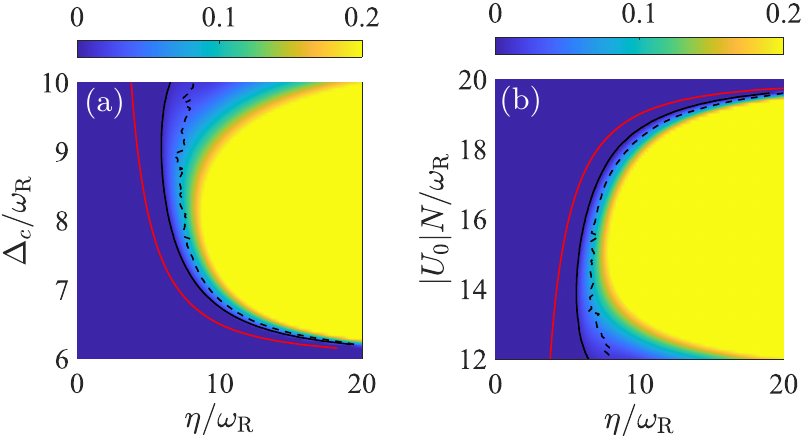}
\caption{Growing rate of the maximally growing mode, with parameters same as that used in the main text Fig.~\ref{fig:phase_diagram}a for (a)  and Fig.~\ref{fig:phase_diagram}b for (b). The black solid lines separate the stable region (left side with zero growing rate) and unstable region (right side with positive growing rate).}
\label{fig:grow_rateA}
\end{figure}

\begin{figure}[b]
\includegraphics[width=1.0 \linewidth]{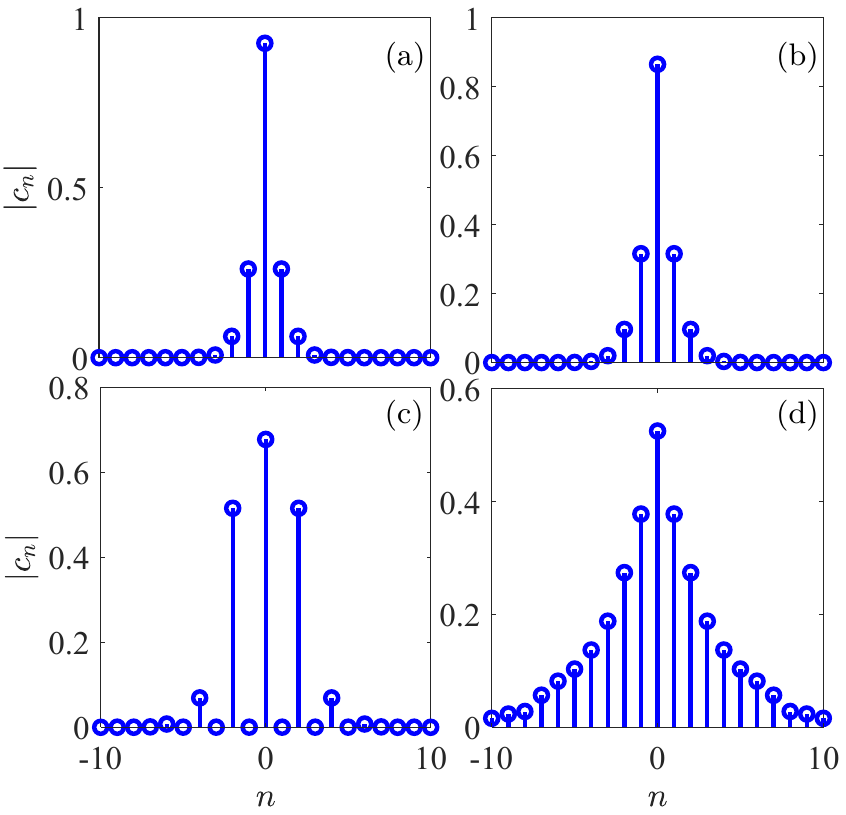}
\caption{Time-averaged amplitudes $|c_n|$ of momentum-space distribution for (a) $\eta=5.2\omega_R$ (S phase), (b) $\eta=6.4\omega_R$(L phase), (c) $\eta=8.8\omega_R$ (AL phase) and (d) $\eta=14\omega_R$ (C phase), with $\Delta_{c}=9$, $U_{0}N  =-12\omega_R $ and $g_\text{aa}=0$.}
\label{fig:BEC_kspaceA}
\end{figure}
 
To analyze the stability of the steady state,
we work in the momentum space 
$\Psi(x,t)=\sqrt{N/L}\sum_n c_n(t) e^{ink_cx}$ and rewrite the equation of motion in the $c_n$ basis
\begin{eqnarray}
  i \partial_t C &=& M C \label{eq:eomcn} \\
  i \partial_t \alpha &=&\left[\Delta_{c}-i \kappa +U_{0} N \mathcal{B}\right] \alpha+\eta\sqrt{N}\Theta.
  \label{eq:eomA2}
\end{eqnarray}
Where $C=[\cdots,c_n,c_{n+1}\cdots]^T$ and $M=(n^2\omega_R+ \frac{U_0}{2}|\alpha|^2) \delta^{(0)}+  \frac{U_0}{4}|\alpha|^2 \delta^{(2)}+ \frac{\eta} {\sqrt{N}} (\alpha+\alpha^*) \delta^{(1)}$, $\mathcal{B}=C^\dag (\frac{\delta^{(0)}}{2}  + \frac{\delta^{(2)}}{4} )C$ and $\Theta=C^\dag\frac{\delta^{(1)}}{2}C$. The matrix $\delta^{(j)}$ has elements
$[\delta^{(j)}]_{n,n'}=\delta_{|n-n'|,j}$.
The steady state solution then can be obtained from Eqs.~\ref{eq:eomcn} and \ref{eq:eomA2}, using the the aforementioned imaginary time evolution method.

The stability of 
these solutions can be determined by
linearizing Eqs.~\ref{eq:eomcn} and \ref{eq:eomA2} around the steady state. We consider the quantum fluctuation on top of the steady state $c_n\rightarrow c_n+\delta c_n$ and $\alpha\rightarrow \alpha +\delta \alpha$, substitute them into Eqs.~\ref{eq:eomcn} and \ref{eq:eomA2} and keep terms up to the first order of the fluctuation, we obtain
\begin{equation}
		i\partial_t\delta \psi=S \delta\psi
\end{equation}
with $\delta\psi=\left[\cdots,\delta c_{n},\cdots, \delta c_{n}^{\dag}, \cdots, \frac{\delta \alpha}{\sqrt{N}}, \frac{\delta \alpha^{\dag}}{\sqrt{N}}\right]^T$ and 
\begin{eqnarray}
	    S= \left[\begin{array}{ccc}
	    S^{11} & 0 & S^{13} \\
		0 & S^{22} & S^{23} \\
		S^{31} & S^{32} & S^{33}
	    \end{array}
	 \right].
\end{eqnarray} 
Where $S^{11}=M-\mu$, $S^{22}=-M^{*}+\mu$, $S^{13}= {[Q^* C}, Q C]$, $S^{23}=[-Q^*C,-QC^*]$, and 
\begin{eqnarray}
    S^{31}&=&\left[\begin{array}{cc} C^\dag Q \\ -C^\dag Q^*\end{array}\right], \text{ } S^{32}=\left[\begin{array}{cc} C^T Q \\ -C^T Q^*\end{array}\right], \nonumber \\
    S^{33}&=&\left[\begin{array}{cc} \Delta_c-i\kappa+U_0N\mathcal{B}  & 0\\ 0&-\Delta_c-i\kappa-U_0N\mathcal{B}\end{array}\right] \nonumber 
\end{eqnarray} 
with
$Q=\alpha U_0{N}\left(\frac{\delta^{(0)}}{2}+\frac{\delta^{(2)}}{4}\right)+\frac{\eta}{2}\delta^{(1)}$. The imaginary parts of the eigenvalues
of $S$ correspond to the growing rates of the collective fluctuation, the eigenvalue for the maximally growing mode has the largest imaginary part. In Fig.~\ref{fig:grow_rateA}, we plot the growing rate of the maximally growing mode with parameters same as that in Figs.~\ref{fig:phase_diagram}a and \ref{fig:phase_diagram}b in the main text. The maximal growing rate changes from zero to positive across the transition to instability.

\textbf{\emph{Momentum space distribution.---}}As we discussed in the main text, the BEC is well localized around $n=0$ in the momentum space even in the chaotic phase. Here we show the time averaged momentum space distribution in Fig.~\ref{fig:BEC_kspaceA} for different dynamical phases. We see clearly that strong pump rate in general leads to broader distribution, and in the AL phase, we see that only the even recoil momenta with $n=0,2,4,\cdots$ are occupied.


%

\end{document}